\documentclass[%
 reprint,
superscriptaddress,
 amsmath,amssymb,
 aps,
prl,
]{revtex4-2}

\usepackage{graphicx}%
\usepackage{dcolumn}%
\usepackage{multirow}
\usepackage{amssymb}
\usepackage{comment}
\usepackage{MnSymbol,wasysym}
 \usepackage{setspace}
\usepackage{array}
\usepackage{makecell} 

\usepackage[dvipsnames,table,xcdraw]{xcolor}

\usepackage{hyperref}
\hypersetup{
    colorlinks=true,
    linkcolor=Blue,
    filecolor=Blue,      
    urlcolor=Blue,
    citecolor=blue,
}

\definecolor{Gray}{gray}{0.85}
\newcolumntype{g}{>{\columncolor{Gray}}r}
\newcolumntype{w}{>{\columncolor{white}}r}
\newcolumntype{C}[1]{>{\centering\arraybackslash}m{#1}}

\usepackage{pifont} 

\newcommand{\D}{\mathrm{d}} 

\begin{document}

\title{
Tunable intertwining via collective excitations
}

\author{Andr\'{a}s Szab\'{o}}
\affiliation{Institute for Theoretical Physics, ETH Zurich, 8093 Zurich, Switzerland}
\affiliation{Max Planck Institute for Solid State Research, D-70569 Stuttgart, Germany}

\author{R. Chitra}
\affiliation{Institute for Theoretical Physics, ETH Zurich, 8093 Zurich, Switzerland}

\date{\today}\textit{}

\begin{abstract}
The intertwining of multiple order parameters is a widespread phenomenon in equilibrium condensed matter systems, yet its exploration is often hindered by the complexity of real materials.
Here, we  present  a controlled study of intertwined orders in a minimal and versatile driven-dissipative quantum-engineered platform. We consider  a Bose-Einstein condensate at the intersection of two optical cavities,  realizing two competing copies of a $\mathbb{Z}_2$ symmetry-breaking superradiant phase transition characterized by density wave orders.
Using periodic drives that exploit dynamical symmetry reduction, we show that collective excitations can be harnessed to stabilize a variety of novel intertwined orders.
Going beyond the conventional phenomenology involving Landau orders, we show the emergence of a larger class of out-of-equilibrium intertwined phases, including intertwining of purely time-crystalline orders, as well as between Landau and time crystal orders.
These results should be observable in state of the art experimental setups.
\end{abstract}

\maketitle 

\emph{Introduction}---%
Complex quantum materials have ushered in the development of new paradigms of physics, exploring a confluence of ordered states in correlated electron systems.
Such materials are often characterized by rich internal symmetries, hosting multiple symmetry-breaking channels, where distinct phases can coexist, and even induce novel ordered states.
Foremost among these, are the concepts of intertwined phases characterized by multi-component order parameters and fluctuation driven vestigial phases described by composite orders~\cite{BabaevBook}.
Though these were  first  studied in the context of  high-temperature cuprate superconductors~\cite{Fradkin:Intertwined,Fernandes:IntertwinedVestigial,Berg2009}, they are now actively explored in diverse solid-state material families, such as Kagome systems~\cite{Neupert2022,Wang2023}, heavy fermions \cite{Kim2016,Gu2023}, or transition metal dichalcogenides~\cite{Davis2021,Candelora2025}.

\begin{figure}[t!]
    \centering
    \includegraphics[width=0.47\textwidth]{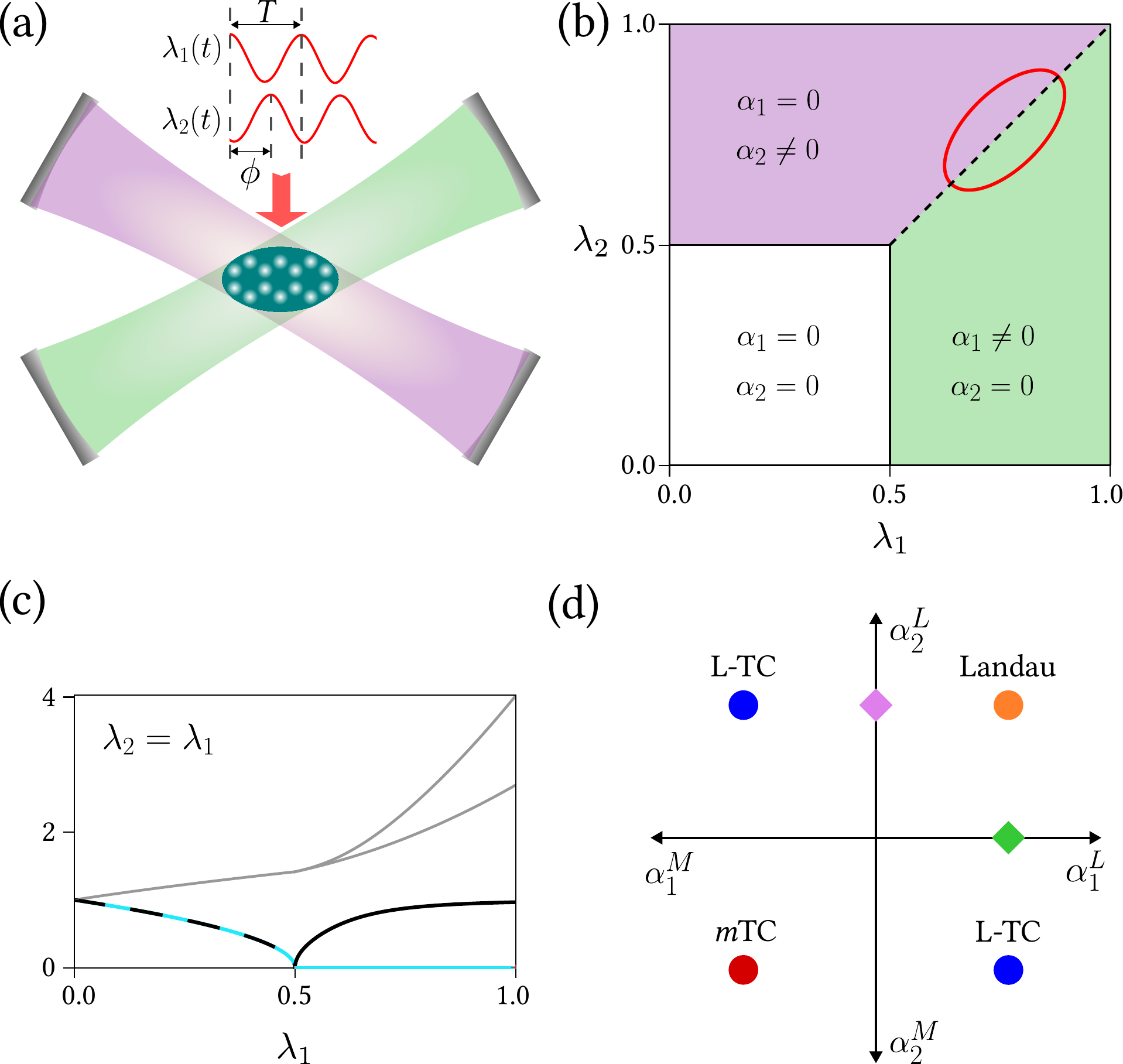}
    \caption{
    (a) Schematic setup of the cross cavity. The BEC (turquoise patch) is illuminated by a transverse laser pump, from which photons scatter into cavity modes 1 and 2 (green and purple).
    The light-matter couplings $\lambda_{1,2}$ are modulated with time period $T$ and a phase difference $\phi$.
    (b) Static phase diagram of the model in Eq.~(\ref{eq:Hamiltonian}) and schematic asynchronous driving scheme (red ellipse). White, green, and purple regions correspond to normal phase, and superradiant phase for cavity 1 and 2, respectively.
    Solid (dashed) lines indicate second (first) order phase transitions.
    The line $\lambda_1=\lambda_2$ is endowed with U(1) symmetry.
    (c) 
    Polariton energies for $\lambda_1=\lambda_2$ 
    , with Goldstone (Higgs) branches shown in cyan (black), and the higher-energy branches in gray.
    (d) Schematic depiction of intertwining, with $\alpha_i^{L,M}$ indicating Landau ($L$) and TC ($M$) order w.r.t. cavity $i$.
    Green (purple) diamonds indicate single-mode superradiance in cavity 1 (2) in the static system.
    }    
    \label{fig:setup_phasediag}
\end{figure}

However, a controlled study of such novel orders in the solid state remains challenging. 
Recently, both classical and quantum light-matter interactions have been proposed as a tool to manipulate competing orders in solid state systems.
Notable examples include photo-induced superconductivity~\cite{Cavalleri2018, Budden2021,de_la_Torre_2021}, floquet engineered quantum matter \cite{Oka_2019}, chiral superconductors \cite{Kitamura_2022} and light-matter intertwined states\cite{kipp2024, Kiffner_2019, Rubio, Schlawin_2022, Eckstein2020, Basov_2025}.
An interesting example  arises in the attractive Hubbard model, where via competition between multiple order channels, a tailored drive  dynamically stabilizes the elusive $\eta$-paired superconducting phase, where the pairs condense at the Brillouin-zone corner momenta\cite{Kitamura2016, Werner2020}.   
Spatially modulated density-wave orders naturally lend themselves to intertwining, as seen for example in the context of pair-density wave superconductors~\cite{Agterberg2020}.  A realm where  matter-wave orders  are rather prevalent is   cavity coupled cold atomic gases.  Prominent examples
include fermionic and bosonic superradiant density-wave orders \cite{Baumann_2010,Baumann_2011, Helson_2023} and supersolidity \cite{Landig_2016, Leonard2017-1}.  As these platforms are  amenable to tailoring  symmetries, they present an opportunity to explore competing orders in a controlled manner.
 Intertwining was demonstrated in BEC-cavity setups~\cite{Demler2017,Morales2018} via the engineering of 
direct interactions that couple different order parameters.   

In this work,  we combine both aspects of coupling to quantum light and driving to show that periodic driving can be harnessed as a tool to realize {a variety of nonequilibrium intertwined phases: intertwined Landau orders, intertwined time crystal (TC) orders and intertwining of Landau and TC orders.}
We explore a minimal model hosting competing order parameters, realized experimentally by placing a BEC at  the intersection of two  crossed cavity modes, each facilitating a $\mathbb{Z}_2$ symmetry-breaking superradiant phase transition  characterized by density wave ordering\cite{Leonard2017-1,Leonard2017-2}.
Threading together the disparate notions of parametric resonance, discrete time crystals\cite{Else_2016,Kessler2021b} and dynamical symmetry reduction leading to massive Goldstone modes, we show that periodic driving of collective excitations  offers a promising route to realizing a variety of novel dynamical phases.

We consider a BEC trapped at the intersection of two crossed  cavity modes, and illuminated by a transverse pump, as depicted in Fig.~\ref{fig:setup_phasediag}(a).  
Photons can scatter from the BEC into either cavity through Raman processes, which leads to the recoil of the atoms into a finite momentum state and simultaneous population of the cavity field, generating a lattice potential for the atoms.
Using a mode expansion which retains the three lowest atomic states, one obtains the following effective low-energy Hamiltonian describing the coupling between the two cavity modes and the  atoms $H=H_1 + H_2$ with \cite{Leonard2017-2}
\begin{align}
    H_{i}=\omega_a a_i^{\dag} a_i + \omega_b b_i^{\dag} b_i + \frac{\lambda_i}{\sqrt{N}} (a_i^{\dag} + a_i) (b_i^{\dag} b_0+b_0^{\dag}b_i) ,\label{eq:Hamiltonian}
\end{align}
where $\omega_a$ and $\omega_b$ are respectively the detuning of the transverse pump relative to the cavity and atomic energies, $\lambda_i$ are the  light-matter couplings, $N$ is the number of atoms, and $a_i,b_i$  denote the annihilation operators for the cavity and the atomic state $i$, respectively.
The atomic number satisfies the constraint  $b_1^{\dag} b_1+ b_2^{\dag} b_2 + b_0^{\dag} b_0=N$.

Each $i=1,2$ on its own describes a simple $\mathbb{Z}_2$-symmetric Dicke model, invariant under the parity operation $P_i(a_i,b_i)=(-a_i,-b_i)$ leading to a 
$\mathbb{Z}_2\times \mathbb{Z}_2$ symmetry for the combined system.
For $\lambda_{1,2} < \lambda_{\rm c}=\sqrt{\omega_a \omega_b}/2$, the system  is in the normal phase (NP), characterized by $\langle a_{1,2} \rangle =0$ and  $\langle b_{1,2} \rangle = 0$.
When one of the couplings $\lambda_i$ crosses the quantum critical point $\lambda_{\mathrm{c}}$, the corresponding parity symmetry is spontaneously broken, and the system enters a superradiant phase (SP) characterized by finite $\langle a_i \rangle$ and $\langle b_i \rangle$.
The two superradiant phases, associated with distinct cavities, are separated by a first-order phase transition.
In standard parlance, $\alpha_1$ and $\alpha_2$ constitute competing order parameters.
We display the phase diagram of the static system in Fig.~\ref{fig:setup_phasediag}(b).

For $\lambda_1=\lambda_2=\lambda_0$ the Hamiltonian is endowed with a U(1) symmetry~\cite{Leonard2017-1}, whereby $(a_1,b_1)$ and $(a_2,b_2)$ can be rotated into each other continuously.
Along the U(1) line, the excitation spectrum of the NP for $\lambda_0<\lambda_{\rm c}$ features two degenerate polariton branches, one of which fully softens at $\lambda_0=\lambda_{\rm c}$ where the U(1) symmetry breaks spontaneously.
This leads to a gapless Goldstone mode and a gapped Higgs mode in the SP, as shown in Fig.~\ref{fig:setup_phasediag}(c), where the associated Landau potential in the $\alpha_1$, $\alpha_2$ space assumes the Mexican hat shape, where $\alpha_{1,2}={\rm Re}\langle a_{1,2}\rangle$.
When the U(1) symmetry is spontaneously broken, both order parameters become nonzero, resulting in a supersolid phase of the bosons, which was observed experimentally~\cite{Leonard2017-1}.
Tuning slightly away from the U(1) line, however, lifts the polariton degeneracy in the NP, and renders the Goldstone mode massive.

\emph{Periodic driving}---%
In the simpler Dicke model, a $\mathbb{Z}_2$-symmetry-preserving drive protocol was shown to result in new dynamical phases that break time translation symmetry~\cite{Chitra2015,Molignini_2018}, whenever the drive was parametrically resonant with a polaritonic mode.
Here, we consider an asynchronous driving protocol in which $\lambda_{1,2}(t)=\lambda_0 + \epsilon \cos(\omega t \mp \phi/2)$ with $\epsilon$ and $\omega$ the driving amplitude and frequency, respectively, and a phase lag $\phi$ between the driven cavities.
When $\phi=0$, we have a fully U(1)-symmetry-preserving drive.
Then, for $\lambda_0 - \epsilon>\lambda_{\rm c}$, the Goldstone mode in the SP remains pinned at zero energy and cannot be resonantly coupled to.
The other polaritonic and Higgs modes can be excited via straightforward parametric resonance akin to the Dicke model.

In contrast, for $\phi\neq 0$ [red trajectory in Fig.~\ref{fig:setup_phasediag}(b)], even though the time-averaged drive preserves the U(1) symmetry, the instantaneous symmetry is $\mathbb{Z}_2 \times \mathbb{Z}_2$.
This implies that all polaritonic excitation modes, including the  original massless Goldstone mode, acquire a fluctuating gap.
This is similar in vein to the drive used to induce $\eta$-paired superconducting states in \cite{Kitamura2016}.
There, the time-averaged drive preserves the full SO(4) symmetry of the system while the instantaneous Hamiltonian typically preserves only the lower SU(2) symmetry.
In our work, a measure of the U(1) symmetry breaking can be obtained by $\Delta(\phi)=\overline{|\lambda_1(t)-\lambda_2(t)|}\sim |\sin(\phi/2)|$, where overline denotes time average and $-\pi \leq \phi \leq \pi$. 
Note however, that  discrete time translation invariance $H(t+nT)=H(t)$ generated by $\mathcal{T}_T^n$ for integer $n$  and $T=\frac{2\pi}{\omega}$ now endows the Hamiltonian with an additional symmetry $\mathbb{Z}^T$.
We use the superscript $T$ to differentiate it from the time independent symmetries.
For $\phi \ne 0$, the system is not amenable to a simple analysis of coupled Mathieu oscillators as in the Dicke model, and necessitates other approaches.

\renewcommand{\arraystretch}{1.13}
\begin{table}[b]
  \centering
  \begin{tabular}{|C{2cm}|C{1.3cm}|C{1.1cm}|C{1.cm}|C{1.cm}|C{1.cm}|C{1.cm}|}
  \hline
  Intertwining & Broken & $\alpha_1^L$ &$\alpha_2^L$   & $\alpha_1^M$  & $\alpha_2^M$\\ 
  \hline
  Landau & $\mathbb{Z}_2 \times \mathbb{Z}_2 $  & $\ne 0$ & $\ne 0$ & $ 0$ &$ 0$ \\
   \emph{m}TC & $\mathbb{Z}_2^T $  & 0 & 0 & $\ne 0$ &$\ne 0$ \\
   L-TC & $\mathbb{Z}_2 \times \mathbb{Z}_2^T $ & $\ne 0$ & 0 & 0&$\ne 0$ \\
    \hline
    \end{tabular}
    \caption{The different intertwined  phases for the two cavity modes  in the  asynchronously  driven system  with $\lambda_0 > \lambda_c$ and $\phi >0$  are summarized.  For the case $\phi < 0$, the same table holds but with the cavity indices permuted.}\label{tab:phases}
\end{table}

\begin{figure*}[!t]
    \centering
    \includegraphics[width=0.95\linewidth]{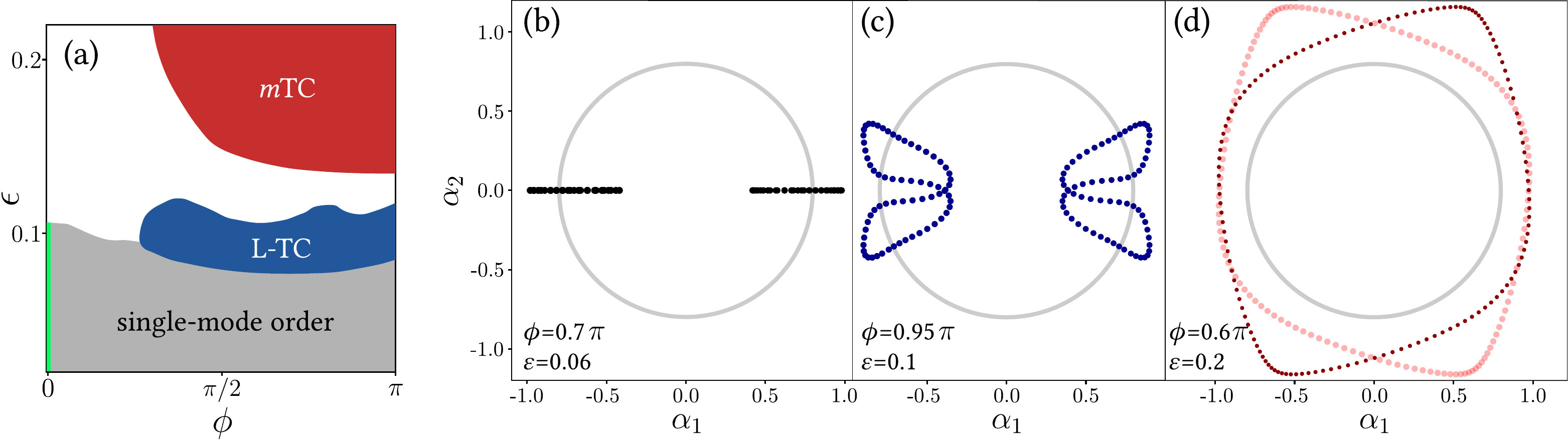}
    \caption{(a) Phase diagram of the Higgs mode in the ($\phi$,$\epsilon$) plane, with $\kappa=0.1$, $\lambda_0=0.85$, and $\omega=0.92$, satisfying parametric resonance with the Higgs polariton branch. All frequencies and units are in terms of $\omega_a$ which has been set equal to one. Here $\phi=0$ is the U(1)-symmetry broken phase, shown in green. While the white region is devoid of a coherent steady state, gray, blue, and red regions indicate periodic symmetry-broken phases, see text. Representative trajectories in the ($\alpha_1$,$\alpha_2$) plane are shown in panels (b), (c), and (d), respectively, where gray circle indicates the static potential minimum. In (d) the two steady state trajectories are shown in dark and pale markers as a guide for the eye. For a complete set of steady states see~\cite{supplementary}.}
    \label{fig:Higgs}
\end{figure*}

\emph{High-frequency expansion}---%
To shed light on the effects of asynchronous driving we employ a high-frequency or Magnus expansion.
To this end, we rewrite the Hamiltonian as
\begin{align}
    H=H_0 +  \epsilon [O_1 \cos(\omega t - \phi/2) +  O_2\cos(\omega t +\phi/2)],
\end{align}
where $H_0=H\rvert_{\epsilon=0}$ contains the time-independent part and $O_i= (a_i + a_i^\dag)(b_i^\dag b_0 + b_0^\dag b_i)/\sqrt{N}$.
A high-frequency  expansion approximates the Floquet Hamiltonian as $H_F=\sum_{n=0}^\infty T^n \Omega_n$, where $T= 2\pi/\omega$ is the period of the drive.
Though this is technically valid only for large driving frequencies (beyond the range of resonant frequencies we are interested in), it is nonetheless illustrative of how $\phi$ generates terms which lead to intertwining.
Here $\Omega_0= H_0$, whereas the first nontrivial term reads
\begin{align}\label{comm}
    2\pi i \hbar  \Omega_1= \epsilon \sin\Big(\frac{\phi}{2}\Big) [H_0,O_1-O_2]+\frac{\epsilon^2}{2}\sin\phi [O_1,O_2],
\end{align}
where the leading-order term in $\epsilon$ is indeed proportional to $\Delta(\phi)$.
Looking at this term, we can separate $H_0=h + \lambda_0 (O_1 + O_2 )$, where $h$ is the noninteracting part of $H_0$, and note that it only involves the commutators  $[h,O_i]$ and $[O_1,O_2]$.
For nonzero phase lag $\phi$, the former yields the terms $(a_i^\dag - a_i)(b_i^\dag b_0+b_0^\dag b_i) $ and $(a_i^\dag + a_i)(b_i^\dag b_0-b_0^\dag b_i)$ which couple to both quadratures of a given cavity mode, akin to the interpolating Dicke-Tavis Cummings model.
The second generates a direct coupling between the two cavities $(a_1^\dag + a_1)(a_2^\dag + a_2)(b_1^\dag b_2 - b_2^\dag b_1)$ mediated by the atoms.
This is reminiscent of the $a_1^\dag a_2+{\rm h.c.}$ term which couples to another atomic excitation mode engineered in~\cite{Morales2018} to induce intertwining.
Already to leading order in $\epsilon$, clearly $\phi\ne 0$ breaks the U(1) symmetry of the Floquet Hamiltonian $H_F$, while preserving $\mathbb{Z}_2 \times \mathbb{Z}_2$, reminiscent of the SO(4)$\to$SU(2) symmetry reduction in the problem of $\eta$-pairing \cite{Kitamura2016}.
This consequently endows the original Goldstone mode with an effective mass without recourse to complex Floquet protocols as done in~\cite{Bukov2024}.

While we here only show the leading-order results of the Magnus expansion in $T$, higher orders involve nested commutators of $h$, $O_1$, and $O_2$, see~\cite{supplementary}, 
highlighting  the potential for intertwined orders.  Caution needs to be exercised regarding the convergence of such high-frequency expansions when one approaches parametric instabilities.
In what follows, using a mean-field approach, we will show that the notion of  intertwining   generalizes in the driven problem to involve intertwining of the usual Landau and discrete time crystalline orders.

\emph{Numerical results}---%
We explore the  stationary states in the presence of cavity dissipation by numerically integrating the equations of motion for the Lindbladian  time evolution of the mean-field expectation values $\langle O \rangle$, where $O= a_i$, $b_i^\dag b_i$, $b_i^\dag b_0$, $b_0^\dag b_0$, $b_1^\dag b_2$, with $i=1,2$.
The mean-field equations are presented in~\cite{supplementary} and are solved numerically for various values of $\epsilon$, $\omega$, and $\phi$.
For simplicity, we set $\omega_a=\omega_b=1$, which in the limit of zero dissipation yields $\lambda_{\rm c}=1/2$. Note that the physical energy scales in the experiment are  in units of  the  atomic recoil energy $E_{\rm r}$ which depends on the atomic mass and the the wave length of the pump laser.
Both cavities are dissipative with a rate $\kappa$, and we only consider driving in the static ordered phases in what follows.

\begin{figure*}[t!]
    \centering
    \includegraphics[width=0.95\linewidth]{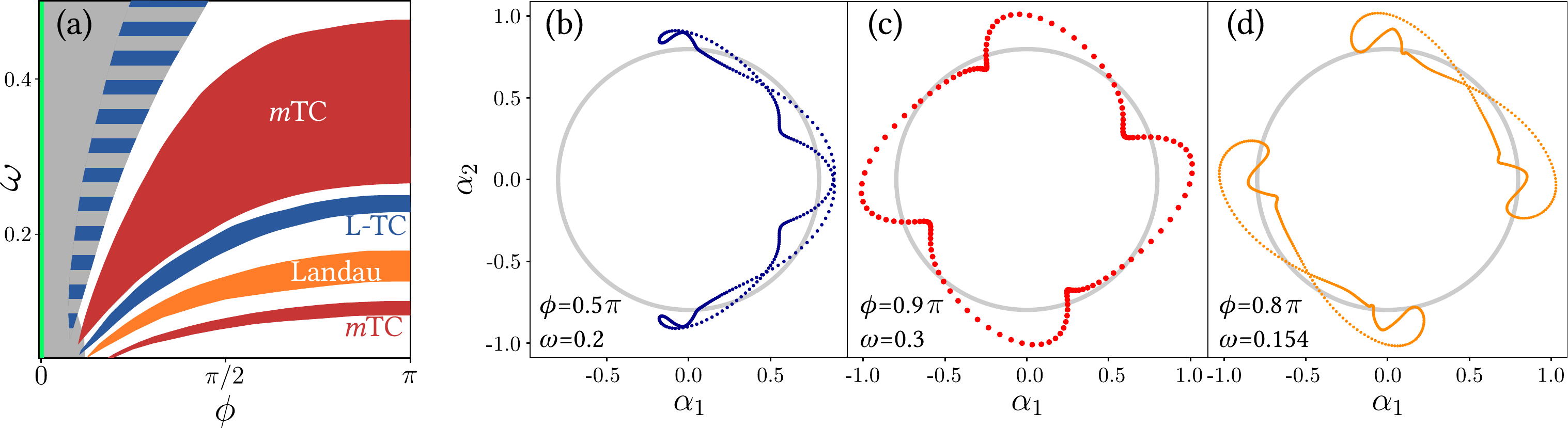}
    \caption{(a) Phase diagram associated with the GS mode in the ($\phi$,$\omega$) plane, with $\epsilon=0.1$, $\lambda_0=0.85$, and $\kappa=0.17$. The white, green and gray regions are the same as in Fig.~\ref{fig:Higgs}. Blue, red, and orange regions indicate symmetry broken phases, with representative steady state trajectories shown in panels (b), (c), and (d),  respectively, see text. Here gray circle shows the static potential minimum.
    The shaded region in (a) indicates coexistence of gray and blue. Highly fine tuned steady states occupying very narrow regions have been removed from the lowest-frequency regime.
    For a complete set of steady states see~\cite{supplementary}.}
    \label{fig:Goldstone}
\end{figure*}

To characterize the plethora of phases obtained, we introduce cavity Landau order parameters via averaging over $n$ drive cycles of period $T$, ${\alpha_i^L}=\int_0^{nT} \alpha_i(t)\, dt$.
As $\alpha_i^L$ is a $\mathbb{Z}_2$ order parameter, it signals two degenerate steady states connected by $P_i$.
In addition to conventional Landau order, the driven system exhibits discrete time-crystalline order, where a mean-field observable $A$ satisfies $\langle A(t+nT)\rangle = \langle A(t)\rangle$ for integer $n>1$.
Focusing on the period-doubling case ($n=2$), we define ${\alpha_i^M} = \tilde{\alpha}_i(\Omega=\frac{\omega}{2})$, with $\tilde{\alpha}_i(\Omega)$ denoting the corresponding Fourier component.
Such a TC also constitutes a $\mathbb{Z}_2$ order, with two degenerate solutions connected by $\mathcal{T}_T\alpha_i(t)=\alpha_i(t+T)=-\alpha_i(t)$.
In what follows, we discuss different intertwined phases that emerge in the driven system, whose characterizing order parameters and  corresponding broken symmetries are summarized in Table~\ref{tab:phases}.


 We first discuss the  case of the  resonantly driven Higgs mode.   
We set $\lambda_0=0.85$ and modulate the drive with frequency $\omega=\Delta_H=0.92$, matching the energy  $\Delta_H$ of the Higgs branch in the static system [see Fig.~\ref{fig:setup_phasediag}(c)].
Note that this is different from the usual parametric driving schemes where one would choose $\omega=2\Delta_H$, as here the relative gap from the Goldstone mode is targeted.
The resulting phase diagram in the $(\phi, \epsilon)$ plane for $0\leq \phi \leq \pi$ (changing the sign of $\phi$ merely exchanges the role of the cavities) is depicted in Fig.~\ref{fig:Higgs}(a).
Here the white region indicates a disordered chaotic phase.
On the other hand, the gray, blue, and red regions describe time periodic steady states with distinct symmetry properties that we will discuss below. Note that these are not the simple Arnold tongues one might expect from standard parametric resonances, rather they stem from a complex interplay of the underlying symmetries.

For comparatively small $\epsilon<\kappa$, we find two $T$ periodic steady states (gray region) with $\alpha_1^L \gtrless 0$, $\alpha_2(t)=0$ (vice versa) for $0\leq\phi\leq \pi$ ($-\pi \leq \phi \leq 0$), representing a single-mode order.
This region has broken $\mathbb{Z}_2$ symmetry and is devoid of intertwining,  see representative trajectories in the $(\alpha_1,\alpha_2)$ plane in Fig.~\ref{fig:Higgs}(b).
When $\epsilon \approx \kappa$, for $|\phi|>\phi_0^{\rm b}\approx 0.3\pi$, we see a bifurcation to the blue regime, see Fig.~\ref{fig:Higgs}(c) for representative trajectories.
Here, for $\phi>0$, the $\alpha_1$ mode shows $\mathbb{Z}_2$ Landau order with $\alpha_1^L\ne 0$ and $T$ periodicity while $\alpha_2$ manifests period-doubled TC order with $\alpha_2^M\ne 0$, but $\alpha_2^L=0$.
This region is characterized by broken $\mathbb{Z}_2\otimes\mathbb{Z}_2^T$ symmetry and four degenerate steady-state solutions are connected by the operations of  $P_1$ and $\mathcal{T}_T$, see extended data in~\cite{supplementary}.
Note that the roles of $\alpha_{1,2}$ are switched when $-\pi<\phi<0$ or $\pi<\phi<2 \pi$.
Curiously, these states exist as metastable solutions when initial conditions are randomly sampled.
The energy levels of the four solutions cross at $\phi=\pi$, where all four states are equally probable, due to the emergent symmetry  where the Hamiltonian is invariant under the composite operation of  $(a_1, b_1)\leftrightarrow (a_2,b_2)$ and $t\to t\pm T/2$.
The blue region, therefore, realizes a novel phase of matter, which displays dynamically intertwined \emph{Landau-time crystal} (L-TC) order.

A further bifurcation to the red region emerges at  $\epsilon>\kappa$ and $|\phi|>\phi_0^{\rm r}$ ($\epsilon$ dependent), see representative trajectories in Fig.~\ref{fig:Higgs}(d), where both cavity modes manifest period doubling, whilst maintaining $\alpha_{1,2}^L=0$.
Here the $\mathbb{Z}_2^T$ symmetry is broken and we obtain four steady states corresponding to a dynamically intertwined multicomponent TC (\emph{m}TC) order parameter $( \alpha_1^M, \alpha_2^M)$.
The four solutions are related by the action of the shift operator $\mathcal{T}_T \alpha_i(t)=-\alpha_i(t)$ for each cavity mode, see~\cite{supplementary}.
Coincidentally, this is equivalent to acting with $P_i$, nevertheless the underlying symmetry breaking is fully in the time domain.
To summarize, driving the Higgs mode can lead to dynamically interwined TC phases where one or both cavity modes break discrete time translation symmetry.

We now explore the regime dominated by the Goldstone mode with a drive-induced dynamical gap.
A measure of this gap is given by  $\Delta(\phi)\propto|\sin(\phi/2)|$ which characterizes the departure from $U(1)$ symmetry.
Driving these modes, we unearth a rich representative steady-state phase diagram in the $(\phi, \omega)$ plane, see Fig.~\ref{fig:Goldstone}(a).
As in the Higgs case, we here highlight representative steady-state trajectories for each intertwined phase, and show a full set of stationary states in~\cite{supplementary}.
While for small $\phi$ we find the same phase shown in gray as in Fig.~\ref{fig:Higgs}, for $\phi > \phi_0^{\rm GS}\sim 0.1 \pi$ we find a  hierarchy of phases with shapes similar to the curve $\Delta(\phi)$.
The threshold $\phi_0^{\rm GS}$ increases for larger $\kappa$, reminiscent of  the shifts in  Arnold lobes  in the presence of dissipation\cite{Zerbe1994}.
Note that the associated trajectories in the $(\alpha_1,\alpha_2)$ plane remain in the vicinity of the static potential minimum, highlighting the low-energy nature of the underlying GS mode.

The blue and red regions (representative trajectories shown in Fig.~\ref{fig:Goldstone}(b) and (c), respectively) are respectively counterparts to the dynamically intertwined L-TC and \emph{m}TC orders for resonant Higgs driving, featuring four steady states via $\mathbb{Z}_2\otimes \mathbb{Z}_2^T $ and $\mathbb{Z}_2^T$ symmetry breaking.
As before, the  $\phi>0$  blue region is characterized by finite $\alpha_1^L$ and $\alpha_2^M$ and features a second set of degenerate metastable states, with the roles of $\alpha_{1,2}$ switched along with a level crossing at $\phi=\pi$.

Surprisingly, a new regime appears in the orange arc in Fig.~\ref{fig:Goldstone}(a) with distinct symmetry properties, driven by ${\rm Im}\langle a_{1,2}\rangle $ and ${\rm Re}\langle b_1^\dag b_2\rangle$. 
It features four degenerate steady states with $\alpha_{1,2}^L\neq 0$ and $\alpha_{1,2}^M=0$, related by parity $P_{1,2}$ for any viable value of $\phi$ [two trajectories shown in Fig.~\ref{fig:Goldstone}(d)].
This indicates the spontaneous breaking of the  $\mathbb{Z}_2 \times \mathbb{Z}_2$ symmetry while preserving the  full $ \mathbb{Z}^T$ time translation symmetry.  This shows that driving can stabilize intertwined Landau orders.
Unlike the  dynamically intertwined  phases, which manifest even when $\lambda_0^{(1)}\ne\lambda_0^{(2)}$, this intertwining of Landau orders requires $\lambda_0^{(1)}= \lambda_0^{(2)}$.
It has its origin in the dynamical symmetry reduction breaking U(1) symmetry instantenously, while leaving $\mathbb{Z}_2 \times \mathbb{Z}_2$ intact, reminiscent of the SO(4)$\to$SU(2) reduction necessary for stabilizing the unconventional $\eta$-pairing.

\emph{Discussion \& outlook}--
Our results can be verified in the well studied cross-cavity experimental setup, where controlled driving via probe fields on the cavities was already used to identify the collective Higgs and Goldstone modes~\cite{LeonardThesis}.
We remark that these novel  intertwined phases exist in the experimentally relevant
parameter regime $\omega_b \ll \omega_a$, see \cite{supplementary}. 
The asynchronous driving can be implemented using a two-color pump, facilitating the independent modulation of the cavity modes. Alternatively, we expect that the low-energy Goldstone mode is accessible via low-frequency modulation of the cavity geometries using piezo-electric devices. The different intertwined phases should be detectable via emitted cavity photons \cite{Kessler2021}.

To conclude, we showed that an interplay between drive and symmetry can be used to generate novel out of equilibrium  intertwined ordered phases in a minimal setup featuring two $\mathbb{Z}_2$ density-wave orders in a tunable cavity platform.
We see that resonant driving of both the Goldstone and massive polaritonic excitations leads to the  intertwining of Landau and/or time crystal order parameters.
It would be exciting to explore if this latter feature involving engineered drives exploring symmetry-enhanced parameter regimes and soft modes provides a general pathway to realizing intertwined and vestigial phases in systems with competing orders.
We leave this for future work.

\begin{acknowledgments}
\emph{Acknowledgement-}
We thank R. Lin for support during the early stages of this work.
We also thank D. Ortu\~no, J. Stefaniak, and M. Bukov for useful discussions.
A.S. is grateful for financial support from the Swiss National Science Foundation (SNSF) through Division II (No. 184739).
\end{acknowledgments}

\section{Supplementary Material}

\subsection{Mean-field theory}
We first discuss the mean-field phase diagram of the closed system.
To obtain the mean-field expectation values of the cavity ($a_i$) and atomic ($b_i$) operators we substitute $a_i = \alpha_i + \hat{a}_i$, $b_i = \beta_i + \hat{b}_i$, and $b_0 = \sqrt{N-b_1^\dag b_1 - b_2^\dag b_2}$ into Eq.~1 of the main text, where Greek letters and capped operators respectively denote mean-field expectation values and fluctuations.
Note that in the closed system $\alpha_i,\beta_i \sim \sqrt{N}$ are real, and we keep only the terms proportional to $N$ to obtain the energy functional
\begin{align}
    E[\alpha_i,\beta_i]= &\sum_{i-1,2} \bigg( \omega_a \alpha_i^2 + \omega_b \beta_i^2 \\ 
    &+4\frac{\lambda_i}{\sqrt{N}}\alpha_i \beta_i
    \sqrt{N-|\beta_1|^2-|\beta_2|^2} \bigg). \nonumber
\end{align}
Minimization with respect to $\alpha_i$ yields
\begin{align}
    \alpha_i=-\frac{2\lambda_i}{\omega_a \sqrt{N}} \beta_i \sqrt{N-\beta^2},
\end{align}
where $\beta^2\equiv \beta_1^2 + \beta_2^2 $.
Substituting the above expression into $E[\alpha_i,\beta_i]$ and minimizing with respect to  $\beta_i$ leads to the equations
\begin{align}
    \omega_b \beta_1-4 \lambda_1^2 \frac{N-\beta^2}{\omega_a N}\beta_1 + \frac{4 \beta_1}{\omega_a N} \sum_i \lambda_i^2 \beta_i^2&=0, \nonumber  \\
    \omega_b \beta_2-4 \lambda_2^2 \frac{N-\beta^2}{\omega_a N}\beta_2 + \frac{4 \beta_2}{\omega_a N} \sum_i \lambda_i^2 \beta_i^2&=0. \label{eq:betamin}
\end{align}

\noindent
{\bf U(1)-symmetric case:} Along the U(1)-symmetric line $\lambda_1=\lambda_2\equiv \lambda$, and the two equations collapse into one and we obtain a solution for $\beta^2$ as
\begin{align}
\beta^2= \frac{N}{2}(1-\mu),
\end{align}
with  $\mu=\frac{\omega_a \omega_b}{4 \lambda^2}$.
The solution for the expectation values therefore spans a U(1) manifold as
\begin{align}
    \beta_1 &= \beta \cos\theta,          & \alpha_1 &= \alpha \cos \theta, \nonumber \\
    \beta_2 &= \beta \sin\theta,          & \alpha_2 &= \alpha \sin \theta, \nonumber \\
    \beta &= \pm \sqrt{\frac{N}{2}(1-\mu)}, & \alpha &= \mp \frac{\lambda }{\omega_a} \sqrt{N (1-\mu^2)}. 
\end{align}

\vspace{0.5cm}
\noindent
{\bf $\mathbb{Z}_2 \times \mathbb{Z}_2$-symmetric case:}
Without the loss of generality, we set $\lambda_2>\lambda_1$ (and $\lambda_2>\lambda_c$), then $\beta_2\neq 0$, but $\beta_1 =0$.
The first equation in (\ref{eq:betamin}) in this case disappears, while in the second we can divide by $\beta_2$ to obtain
\begin{align}
    \beta_1 &= 0,          & \alpha_1 &= 0, \nonumber\\
    \beta_2 &= \pm \sqrt{\frac{N}{2}(1-\mu_2)}, & \alpha_2 &= \mp \frac{\lambda }{\omega_a} \sqrt{N (1-\mu_2^2)},
\end{align}
with $\mu_i=\frac{\omega_a \omega_b}{4 \lambda_i^2}$.
The solution for $\lambda_1>\lambda_2$ follows in an identical way, and these results lead to the phase diagram indicated in Fig.~1(b) in the main body of the paper.

\subsection{Fluctuations}
To calculate the polariton energies we first construct the Hamiltonian describing fluctuations around the mean-field expectation values.
We once more make the substitution $a_i = \alpha_i + \hat{a}_i$, $b_i = \beta_i + \hat{b}_i$ into Eq.~1 of the main text, supplying $\alpha_i$ and $\beta_i$ from the previous section, then collect the terms quadratic in the cavity and atomic operators.
The fluctuation Hamiltonian $H_F$ takes the general form
\begin{align}
    H_F= \omega_a a^\dag a + \omega_b b^\dag b + \lambda (a^\dag + a)(b^\dag + b) + \rho (b^\dag + b)^2.
\end{align}
Defining $\tilde{\omega}_b^2=\omega_b^2 + 4 \omega_b \rho$ and $\tilde{\lambda}=\lambda \sqrt{\omega_b /\tilde{\omega}_b}$, normal mode transformation yields
\begin{align}
    \epsilon_{\pm}^2= \frac{1}{2}\bigg( \omega_a^2 + \tilde{\omega}_b^2 \pm \sqrt{(\omega_a^2 - \tilde{\omega}_b^2)^2 + 16 \tilde{\lambda}^2 \omega_a \tilde{\omega}_b} \bigg).\label{eq:normalmodes}
\end{align}
The polariton energies in the normal phase are obtained via the substitution $\omega_a \to 1$, $\omega_b\to 1$, $\lambda\to \lambda_i$, $\rho \to 0$ in Eq.~(\ref{eq:normalmodes}), yielding
\begin{align}
    \big(\epsilon_{\pm}^{\rm NP}\big)^2=1\pm 2\lambda_i.
\end{align}
Once again we address the superradiant phase in the presence and absence of U(1) symmetry separately.

\vspace{0.5cm}
\noindent
{\bf U(1)-symmetric case:} To show the emergence of Goldstone and Higgs modes, we recapitulate the results from Ref.~\cite{LeonardThesis}.
We make a change of variables in $H_F$ as
\begin{align*}
    b_H&=\cos\theta b_1 + \sin\theta b_2,  &  b_1&=-\sin\theta b_G + \cos\theta b_H, \\
    b_G&=-\sin\theta b_1 + \cos\theta b_2, &  b_2&= \cos\theta b_G + \sin\theta b_H,
\end{align*}
after which the Hamiltonian falls apart into Goldstone and Higgs contributions, $H[a_i,b_i]=H_G + H_H$, where
\allowdisplaybreaks
\begin{align}
    H_G&= \omega_a a_G^\dag a_G + \omega_b \frac{1+\mu}{2\mu} b_G^\dag b_G  \nonumber \\
    &+\lambda\sqrt{\frac{1+\mu}{2}} (a_G^\dag + a_G)(b_G^\dag + b_G), \nonumber\\
    H_H&= \omega_a a_H^\dag a_H + \omega_b \frac{1+\mu}{2\mu} b_H^\dag b_H \nonumber \\
     &+\lambda\mu\sqrt{\frac{2}{1+\mu}} (a_H^\dag + a_H)(b_H^\dag + b_H) \nonumber \\
    &+\omega_b \frac{(1-\mu)(3+\mu)}{8\mu(1+\mu)}(b_H^\dag + b_H)^2.
\end{align}
Reading off $\tilde{\omega}_b$, $\tilde{\lambda}$, and $\rho$ yields the polariton branches as
\begin{align}
\big(\epsilon^{\rm GS}_{\pm}\big)^2&= \frac{1}{2} \Bigg\{ 1+ \Big( \frac{1+\mu}{2\mu} \Big)^2 \nonumber \\
    &\pm \sqrt{\bigg[ 1-\Big( \frac{1+\mu}{2\mu} \Big)^2\bigg]^2 +4 \lambda^2 \frac{(1+\mu)^2}{\mu}}\Bigg\}, \nonumber \\
\big(\epsilon^{\rm H}_{\pm}\big)^2&= \frac{1}{2} \Bigg\{ 1+\frac{1}{\mu^2} \pm \sqrt{\frac{5 \mu^4 - 2\mu^2 + 1}{\mu^4}}\Bigg\}.
\end{align}

\vspace{0.5cm}
\noindent
{\bf $\mathbb{Z}_2 \times \mathbb{Z}_2$-symmetric case:}Just like before we consider $\lambda_2>\lambda_1$ (and $\lambda_2>\lambda_c$), then $\alpha_2,\beta_2\neq 0$, but $\alpha_1, \beta_1 =0$.
We supply the corresponding mean-field expectation values and notice that the fluctuation Hamiltonian again falls apart into contributions from cavity 1 and cavity 2 as
\allowdisplaybreaks
\begin{align}
    H_1&= \omega_a a_1^\dag a_1 + \omega_b \frac{1+\mu_2}{2\mu_2} b_1^\dag b_1 \nonumber \\
    &+ \lambda_1\sqrt{\frac{1+\mu_2}{2}} (a_1^\dag + a_1)(b_1^\dag + b_1),\label{eq:gapped_Goldstone} \\
    H_2&= \omega_a a_2^\dag a_2 + \omega_b \frac{1+\mu_2}{2\mu_2} b_2^\dag b_2 \nonumber \\
    &+ \lambda_2\mu_2\sqrt{\frac{2}{1+\mu_2}} (a_2^\dag + a_2)(b_2^\dag + b_2) \nonumber \\
    &+\omega_b \frac{(1-\mu_2)(3+\mu_2)}{8\mu_2(1+\mu_2)}(b_2^\dag + b_2)^2,
\end{align}
ultimately resulting in the normal modes
\allowdisplaybreaks
\begin{align}
    \big(\epsilon^1_{\pm}\big)^2= \frac{1}{2} \Bigg\{ 1&+ \Big( \frac{1+\mu_2}{2\mu_2} \Big)^2 \nonumber \\
    &\pm \sqrt{\bigg[ 1-\Big( \frac{1+\mu_2}{2\mu_2} \Big)^2\bigg]^2 +4 \lambda_1^2 \frac{(1+\mu_2)^2}{\mu_2}}\Bigg\}, \nonumber \\
    \big(\epsilon^2_{\pm}\big)^2= \frac{1}{2} \Bigg\{ 1&+\frac{1}{\mu_2^2} \pm \sqrt{\frac{5 \mu_2^4 - 2\mu_2^2 + 1}{\mu_2^4}}\Bigg\}.
\end{align}
We show the polariton energies in the NP and SR phases in Fig.~1(c) of the main text.

\subsection{High-frequency expansion}
The Floquet Hamiltonian is given by
\begin{equation}
H_F= \frac{-i \hbar}{T} \ln(1+\sum_{n=1}^\infty P_n) \equiv \sum_{n=0}^\infty T^n \Omega_n,
\end{equation}
where the operators $P_i$ are defined as
\begin{equation}
P_n= \Big(\frac{-i}{\hbar}\Big)^n\int_0^T dt_1\ \dots \int_0^{t_{n-1}} dt_n H(t_1)\dots H(t_n).
\end{equation}
In the present problem, we first rewrite the time-dependent Hamiltonian as
\begin{equation}
H(t)= H_0 +  \epsilon O_1 \cos\left(\omega t - {\frac{\phi}2}\right) +  \epsilon O_2 \cos\left(\omega  t + {\frac{\phi}2}\right),
\end{equation}
where $H_0\equiv  H\rvert_{\epsilon=0}$ and 
\begin{eqnarray}
O_1&=& \frac{1}{\sqrt{N}} (a_1^{\dag} + a_1) (b_1^{\dag} b_0+b_0^{\dag}b_1), \nonumber \\
O_2&=&\frac{1}{\sqrt{N}} (a_2^{\dag} + a_2) (b_2^{\dag} b_0+b_0^{\dag}b_2).
\end{eqnarray}
Performing the nested time integrals, we obtain the following leading order terms in the expansion
\allowdisplaybreaks
\begin{align}
P_1 &= -\frac{i}{\hbar}  T H_0, \nonumber \\
P_2 &=  \Big(-\frac{i}{\hbar}\Big)^2 \frac{T^2}{2\pi} \Big( \pi H_0^2 + \epsilon\sin(\phi/2)  [H_0, O_1-O_2] \nonumber \\
&\ + \frac{\epsilon^2}{2} \sin{\phi} [O_1,O_2]. \Big)
\end{align}
To this order, we can now obtain the leading contribution to the Floquet Hamiltonian in the high driving frequency limit:
\begin{eqnarray}
\Omega_0 &=& H_0,  \nonumber\\
\Omega_1&=&   \frac{i\hbar}{T^2} \bigg( P_2 -\frac{P_1^2}{2}\bigg)  \\
&=& \frac{1}{2 \pi i \hbar} \Big(\epsilon\sin{\frac{\phi}2} [H_0, O_1-O_2] + \frac{\epsilon^2}{2}\sin\phi [O_1,O_2] \Big).\nonumber
\end{eqnarray}

\vspace*{0.5cm}

We first focus on the first nontrivial order $\Omega_1$.
As we are primarily concerned with the role of Higgs and Goldstone excitations, we only consider driving amplitudes $\epsilon/\lambda_0 \ll 1$ in this work.
Then, one can potentially neglect terms of $\mathcal{O}(\epsilon^2)$ and higher in the Magnus expansion.
Looking at the $\mathcal{O}(\epsilon)$ term, we rewrite $H_0$ as
\begin{align}
H_0 &= \sum_{i=1,2}\Big[  \omega_a a_i^{\dag} a_i + \omega_b b_i^{\dag} b_i \Big] +  \lambda_0 (O_1 + O_2) \nonumber\\
   &\equiv h +  \lambda_0(O_1 + O_2), 
\end{align}
such that $[H_0, O_1 - O_2] = [h, O_1 - O_2] - 2 \lambda_0 [O_1,O_2]$.
Calculating the commutators, we find
\allowdisplaybreaks
\begin{align}
&[h,O_1-O_2]  = \nonumber \\
&\frac{\omega_a}{\sqrt{N}} [ 
(a_1^{\dag} - a_1) (b_1^{\dag} b_0+b_0^{\dag}b_1) -
(a_2^{\dag} - a_2) (b_2^{\dag} b_0+b_0^{\dag}b_2) ] \nonumber \\
+& \frac{ \omega_b}{\sqrt{N}} [
(a_1^{\dag} + a_1) (b_1^{\dag} b_0- b_0^{\dag}b_1) -
(a_2^{\dag} + a_2) (b_2^{\dag} b_0-b_0^{\dag}b_2)], \nonumber \\ 
&\lambda_0[O_1,O_2] = {\frac{\lambda_0}{N}} (a_1^{\dag} + a_1)(a_2^{\dag} + a_2) (b_1^{\dag} b_2-b_2^{\dag}b_1).
\end{align}
The first term shows that  inducing a nonzero phase lag between the drives generates terms in the Hamiltonian which now couple to both quadratures of a given cavity mode.
The terms involving the  commutator $[O_1,O_2]$  lead to a direct coupling between the two order parameter sectors.
This is reminiscent of intertwining studied earlier~\cite{Morales2018}.

Going to the next order, $P_3$ is given by
\begin{align}
  (i\hbar)^3P_3 = {\frac{4 \pi^3}{\omega^3}} H_0^3 + \sum_{n=1}^3 \epsilon^n C_n,\label{eq:P3}
\end{align}
with
\begin{widetext}
\begin{align}
    C_1 &=2 \cos{\frac{\phi}2} [H_0,[H_0, O_1+O_2]] +2 \pi \sin{\frac{\phi}2} [H_0^2, O_2-O_1],\nonumber \\
    C_2 &=[[H_0,O_1],O_1]+[[H_0,O_2],O_2]
    -\sin\phi \{H_0,[O_1,O_2]\}
- \Big(\cos\phi-\frac12 \Big) \Big([[O_1,H_0],O_2] + [[O_2,H_0],O_1]\Big),\nonumber\\
   C_3 &=-\sin{\frac{\phi}2}\sin\phi [[O_1,O_2],O_2-O_1].\label{eq:Cn}
\end{align}
\end{widetext}
where curly brackets denote anticommutators.
The second order correction to the Floquet Hamiltonian then takes the form:
\begin{equation}
\Omega_2= \frac{i \hbar}{T^3} \Big(P_3 + {\frac13} P_1^3 - {\frac12}(P_1P_2 +P_2P_1)\Big)
\end{equation}
A simple glance does show that in general the terms in the Floquet Hamiltonian are rather complicated as we increase the order of the expansion.
Clearly, they do not sum to a simple closed form.
However, one can still obtain interesting information from such an expansion in certain limits.
In Eqs.~(\ref{eq:P3}) and (\ref{eq:Cn}), one finds that in first order in $\epsilon$, we typically get terms proportional to   $\cos(\phi/2) $ and   $\sin(\phi/2)$.
On the other hand, in the $\mathcal{O}(\epsilon^2)$ terms the phase lag enters as $\cos\phi$ and $\sin\phi$.
At a given order $n \ge 3$,  the $\cos(\phi/2) $ term involves  a sum of 
$ [H_0^{n-2},[H_0, O_1+O_2]] $ and $H_0\{ O_1+O_2, H_0^{n-3}\} H_0$  alongside  other terms involving nested commutators.
The $\sin(\phi/2)$ terms on the other hand  typically involve $[H_0^{n-1}, O_2-O_1]$.

In the extreme limit where both cavity and atomic detunings are zero, we have $H_0= \lambda_0 (O_1+O_2)$.
Consequently, to generic orders of the Magnus expansion, we have the leading order in $\epsilon$ terms which are proportional to $ \cos(\phi/2)  \sum_n p_n (O_1+O_2)^n +  \sin(\phi/2)\sum_{n}\sum_{m=0}^{n-2} q_{n,m}(O_1+O_2)^{n-m-2} [O_1,O_2] (O_1+O_2)^m$ where $p_n$ and $q_{n,m}$ are numerical prefactors.  In this case, we see that if $\phi=0$, $\rm{Im} \langle b_1^\dag b_2 \rangle= 0$, implying that a non-zero  $\phi$ is indeed necessary to establish this imaginary part, which will be maximal at $\phi=\pi$. Our simplifications highlight that drive indeed generates an intertwining of the different order parameter sectors.  As we show in the main text,  this leads to complex dynamical behaviour.

\subsection{Heisenberg equations of motion}
We write the equations of motion as $\D O/\D t=i \big[ H,O\big]$ describing unitary time evolution.
In case of the cavity operators $a_i$ we incorporate leakage from the cavity as $\D a_i/\D t=i \big[ H,a_i\big]-\kappa a_i$, with $\kappa$ parametrizing dissipation.
The Heisenberg equations of motion read (everything shall be understood in terms of expectation values)
\allowdisplaybreaks
\begin{align*}
    \frac{\D a_i}{\D t} &= -i\bigg[ (\omega_a-i\kappa )a_i-\frac{\lambda_i}{\sqrt{N}}(b_i^{\dag} b_0+b_0^{\dag}b_i) \bigg],\nonumber\\
   \frac{\D b_i^{\dag} b_i }{\D t} &=  i\frac{\lambda_i}{\sqrt{N}}(a_i^{\dag} + a_i)(b_0^{\dag} b_i-b_i^{\dag}b_0), \nonumber\\
    \frac{\D b_0^{\dag} b_0 }{\D t} &=  -i\sum_i \frac{\lambda_i}{\sqrt{N}}(a_i^{\dag} + a_i)(b_0^{\dag} b_i-b_i^{\dag}b_0), \nonumber\\   
    \frac{\D b_1^{\dag} b_0 }{\D t} &= i \bigg[ \omega_b b_1^{\dag} b_0 + \frac{\lambda_1}{\sqrt{N}}(a_1^{\dag} + a_1) (b_0^{\dag} b_0-b_1^{\dag}b_1) \\
    &- \frac{\lambda_2}{\sqrt{N}}(a_2^{\dag} + a_2)b_1^{\dag} b_2 \bigg], \nonumber\\
   \frac{\D b_2^{\dag} b_0 }{\D t} &= i \bigg[ \omega b_2^{\dag} b_0 + \frac{\lambda_2}{\sqrt{N}}(a_2^{\dag} + a_2) (b_0^{\dag} b_0-b_2^{\dag}b_2) \\
   &- \frac{\lambda_1}{\sqrt{N}}(a_1^{\dag} + a_1)b_2^{\dag} b_1 \bigg], \nonumber\\
    \frac{\D b_1^{\dag} b_2 }{\D t} &= i \bigg[ \frac{\lambda_1}{\sqrt{N}} (a_1^{\dag} + a_1) b_0^{\dag}b_2 - \frac{\lambda_2}{\sqrt{N}} (a_2^{\dag} + a_2) b_1^{\dag}b_0 \bigg].
\end{align*}

\begin{figure}[!t]
\vspace{10pt}
    \centering
    \includegraphics[width=0.5\linewidth]{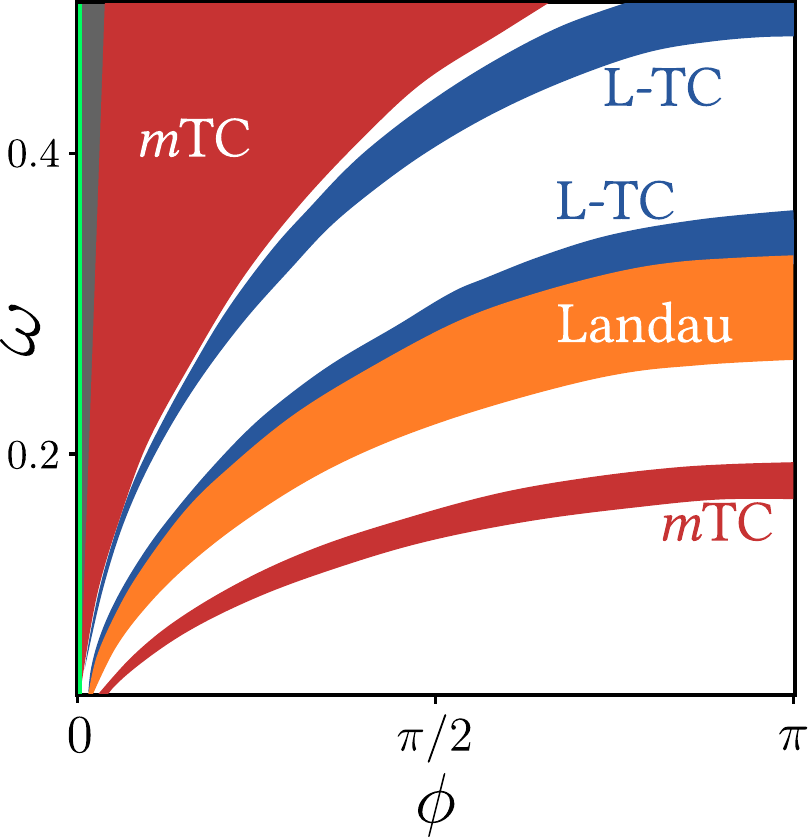}
    \caption{Phase diagram with the GS mode in the $(\phi, \omega)$ plane, with cavity detuning $\omega_a=10$ an order of magntitude larger than the atomic detuning $\omega_b = 1$. The rest of the parameters in untis of $\omega_b$ are $\lambda_0 =2.5$, $\epsilon=0.1$, $\kappa=0.17$. Colors represent the same phases as in Fig.~3 of the main text. Highly fine tuned steady states occupying very narrow regions have been removed from the lowest-frequency regime.}
    \label{fig:Goldstone_q10}
\end{figure}

\section{Extended Data}
Here, we present the phase diagram corresponding to the Goldstone mode for the experimentally relevant regime with dispersively coupled cavity, $\omega_a=10$ an order of magnitude larger than $\omega_b=1$ see Fig.~\ref{fig:Goldstone_q10}.
All intertwined phases are present for these alternative parameters, demonstrating their stability, and can be accessed by tuning $\omega$ and $\phi$.

\begin{figure*}
    \includegraphics[width=11cm]{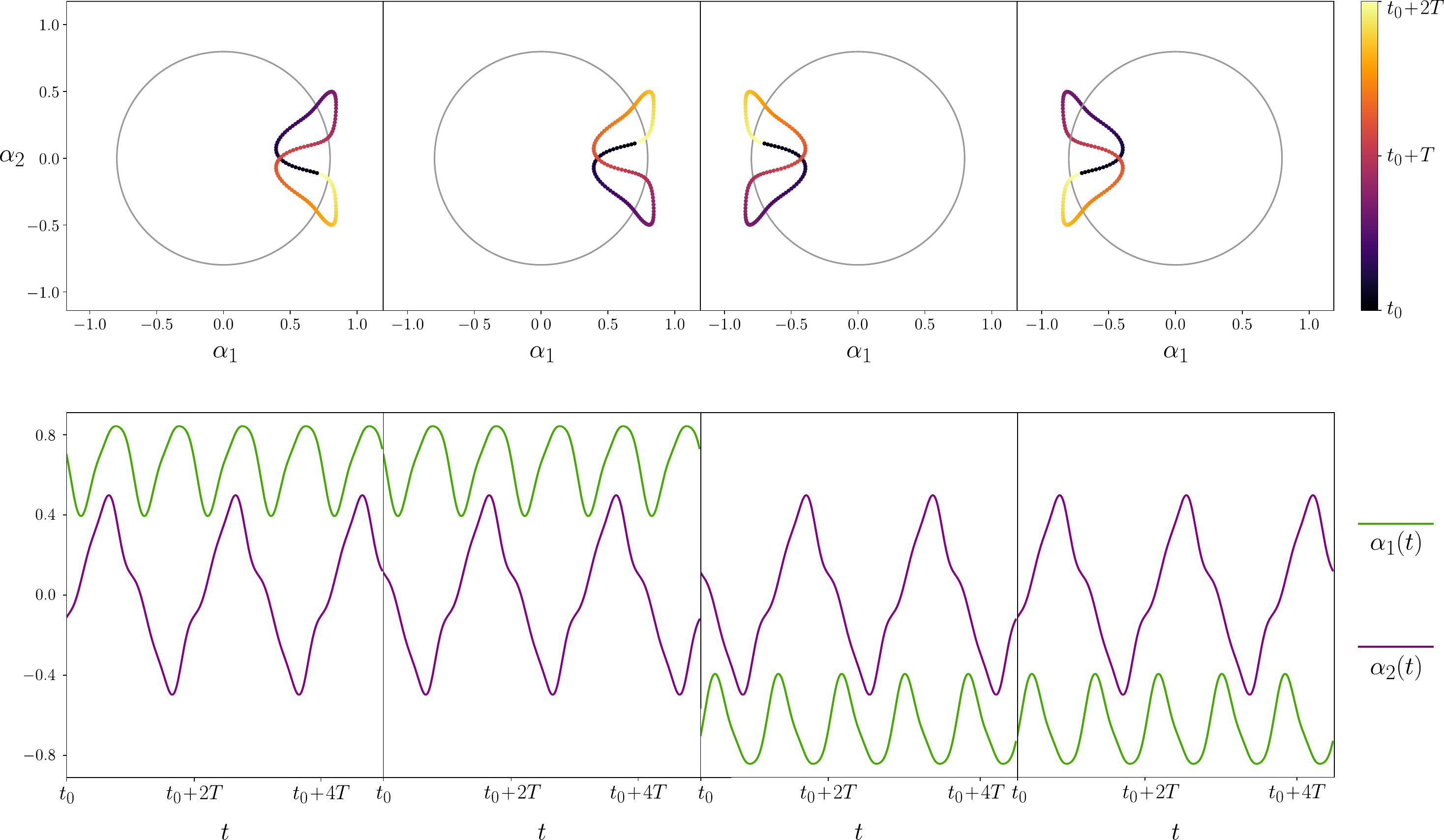}
    \caption{Four steady states corresponding to the L-TC phase for the Higgs resonance with $\lambda=0.85$, $\omega=0.92$, $\epsilon=0.1$, $\phi=0.8$, $\kappa=0.1$.}
    \label{fig:SteadyStates_HLTC}
\end{figure*}

\begin{figure*}
    \includegraphics[width=11cm]{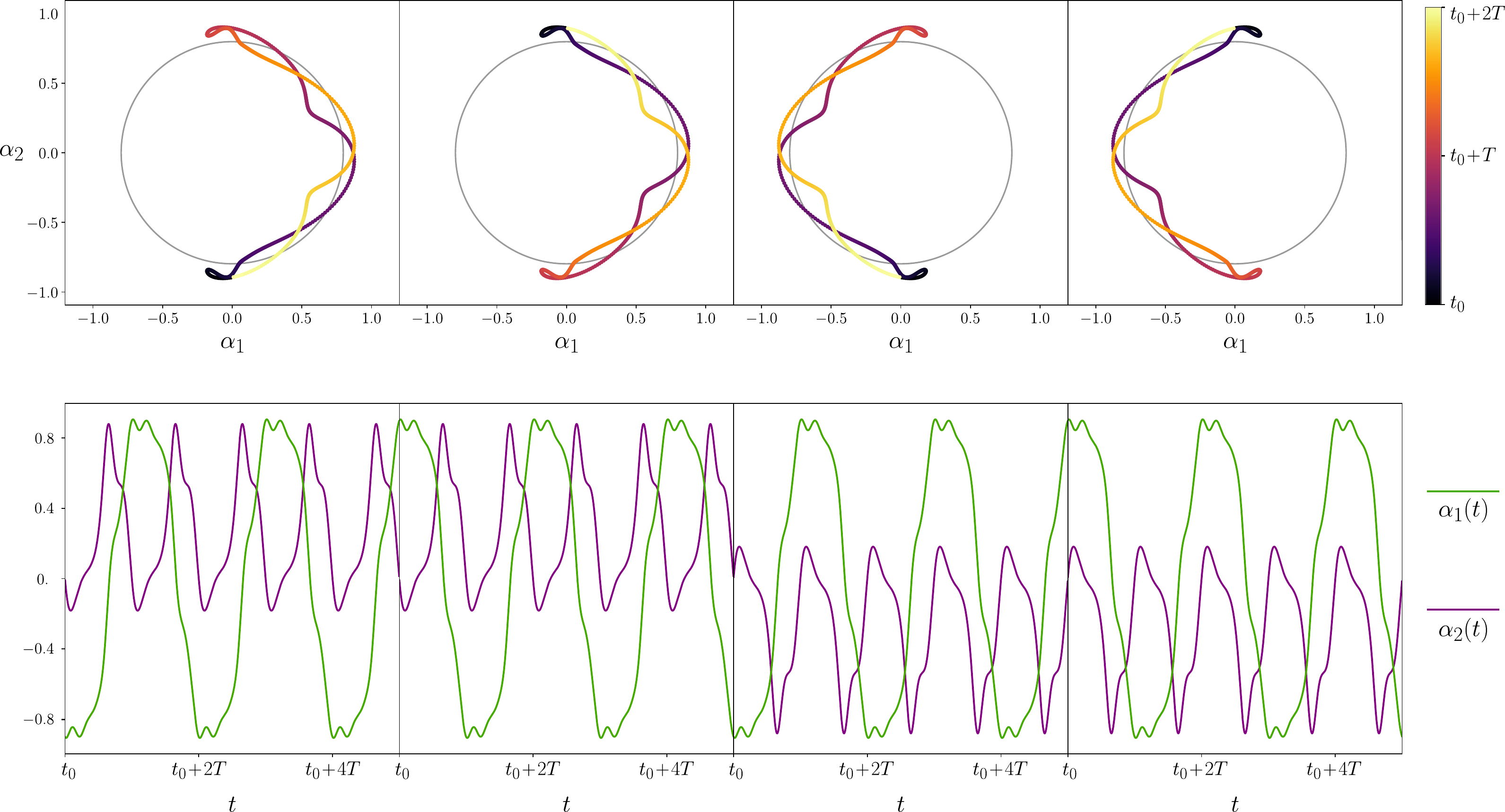}
    \caption{Four steady states corresponding to the L-TC phase for the Goldstone resonance with $\lambda=0.85$, $\omega=0.2$, $\epsilon=0.1$, $\phi=0.5$, $\kappa=0.17$.}
    \label{fig:SteadyStates_GSLTC}
\end{figure*}

Next, we present representative plots of the degenerate steady states in the intertwined phases discussed in the main text.
Figs.~\ref{fig:SteadyStates_HLTC} and \ref{fig:SteadyStates_GSLTC} show four steady states in the intertwined L-TC phase in for the Higgs and Goldstone modes, respectively, connected by applying $P_1$ and $\mathcal{T}_T$.

\begin{figure*}
    \includegraphics[width=11cm]{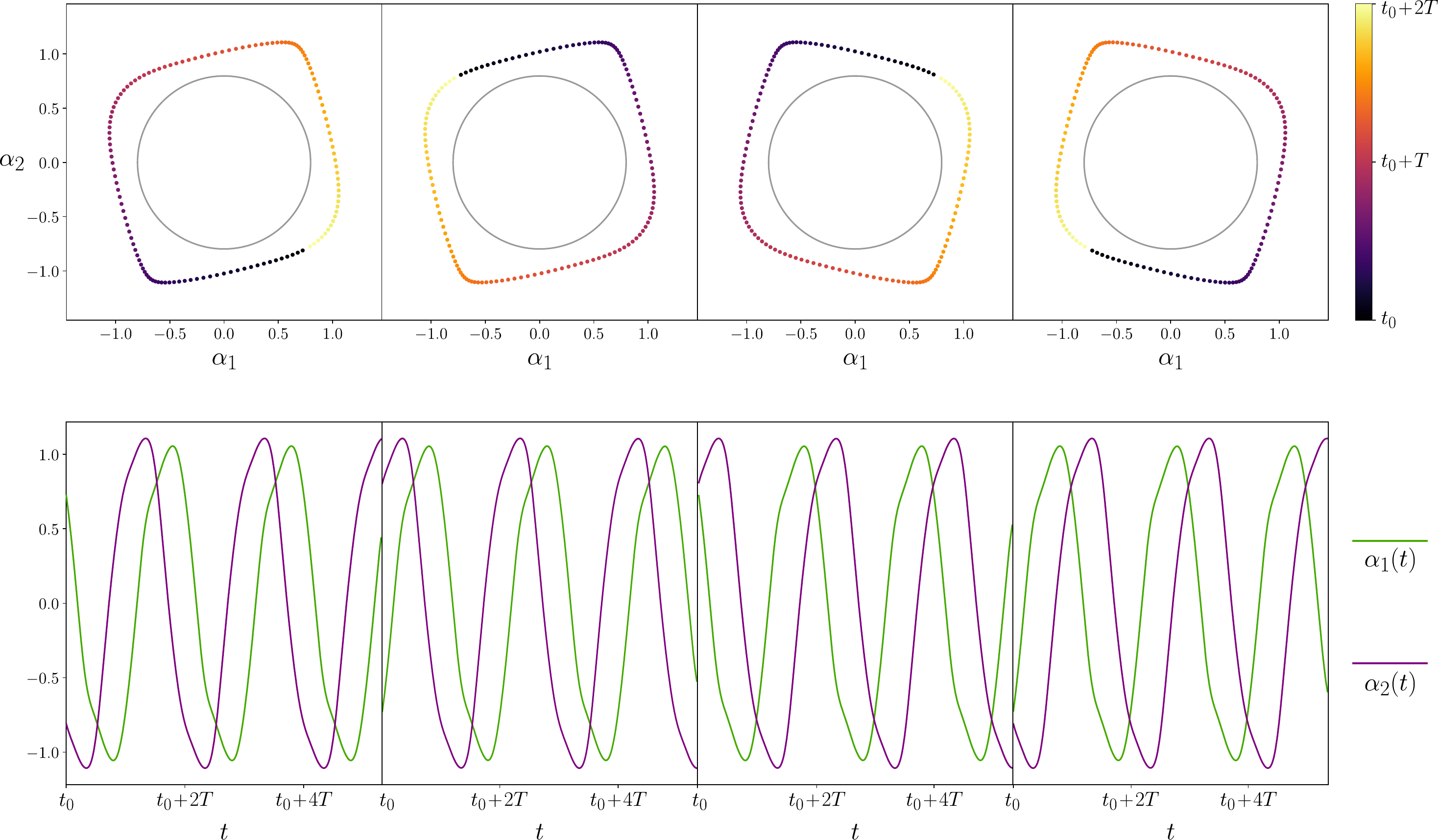}
    \caption{Four steady states corresponding to the MTC phase for the Higgs resonance with $\lambda=0.85$, $\omega=0.92$, $\epsilon=0.2$, $\phi=0.8$, $\kappa=0.1$.}
    \label{fig:SteadyStates_HMTC}
\end{figure*}

\begin{figure*}
    \includegraphics[width=11cm]{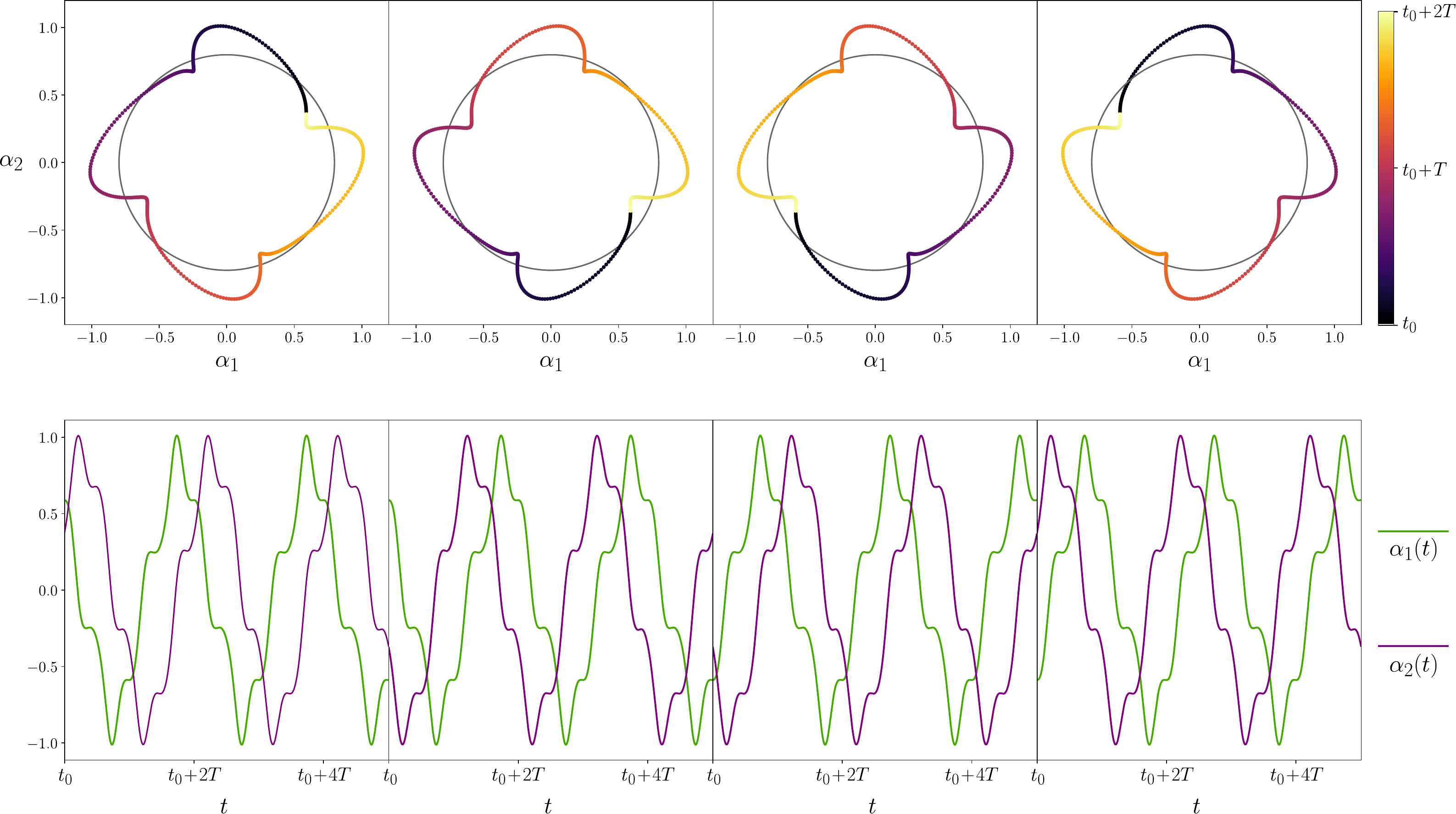}
    \caption{Four steady states corresponding to the L-TC phase for the Goldstone resonance with $\lambda=0.85$, $\omega=0.3$, $\epsilon=0.1$, $\phi=0.9$, $\kappa=0.17$.}
    \label{fig:SteadyStates_GSMTC}
\end{figure*}

In Figs.~\ref{fig:SteadyStates_HMTC} and \ref{fig:SteadyStates_GSMTC} we display the degenerate steady states for the intertwined \emph{m}TC phases respectively for the Higgs and Goldstone modes.
The states are connected by applying $\mathcal{T}_T$ to each cavity mode.

\begin{figure*}
    \includegraphics[width=11cm]{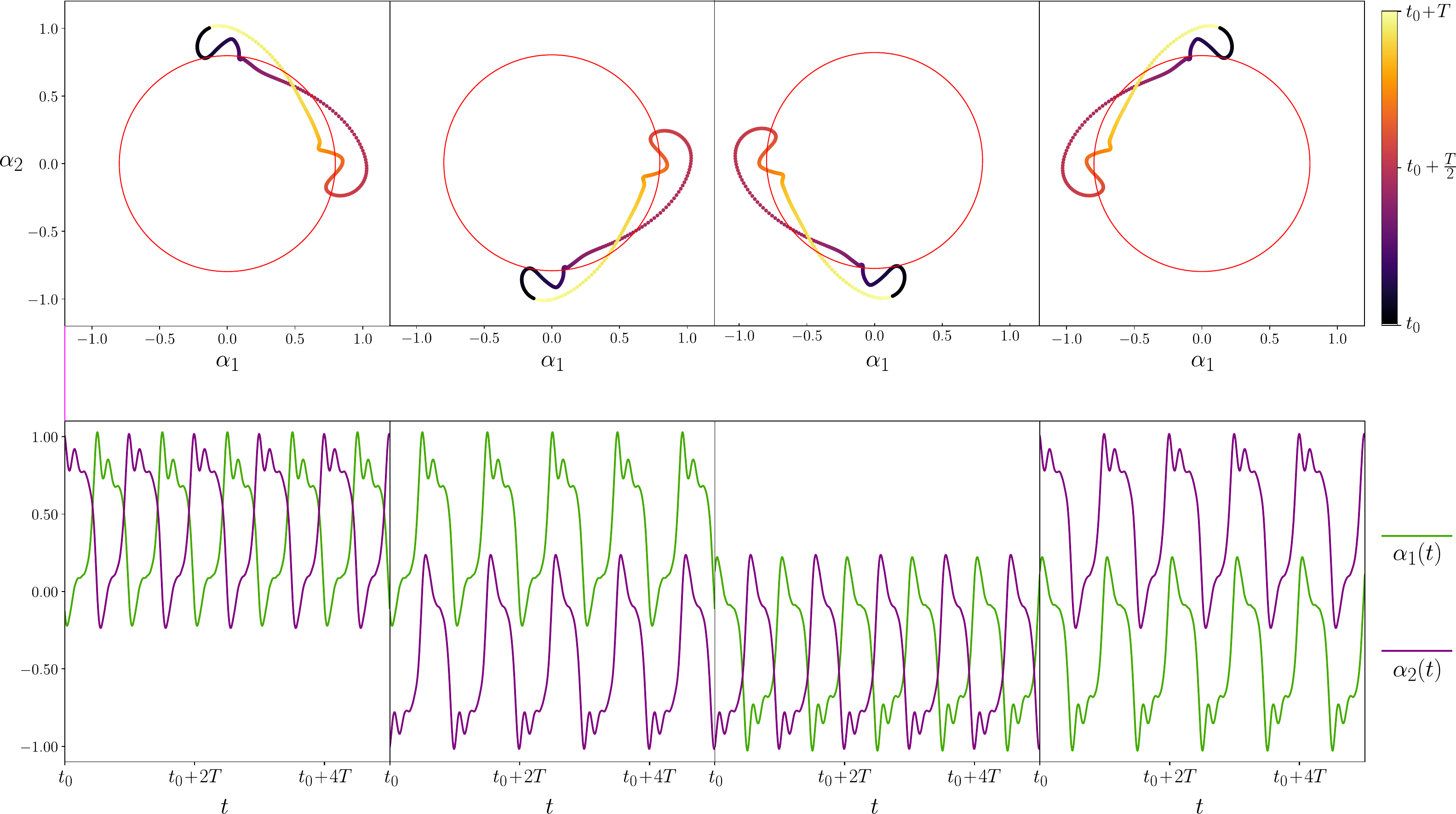}
    \caption{Four steady states corresponding to the Landau intertwined phase for the Goldstone resonance with $\lambda=0.85$, $\omega=0.154$, $\epsilon=0.1$, $\phi=0.8$, $\kappa=0.17$.}
    \label{fig:SteadyStates_GSL}
\end{figure*}

Finally, Fig.~\ref{fig:SteadyStates_GSL} shows four degenerate steady states in the Landau intertwined phase, present in the phase diagram for the Goldstone mode.
Here, the steady states are connected by applying the parity operators $P_{1,2}$.

\bibliography{refs-szabo}
\end{document}